\documentclass[preprint,12pt]{elsarticle}

\usepackage{comment}
\usepackage[all]{nowidow}

\usepackage{rotating}
\usepackage{subfig}

\usepackage{todonotes}

\usepackage{listings}
\usepackage{tabularx}
\usepackage{booktabs}
\usepackage{longtable}

\usepackage{xcolor}

\usepackage{fancyhdr}

\usepackage{amssymb}

\newcounter{bla}

\journal{Computer Physics Communications}

\makeatletter
\newcounter{savesection}
\newcounter{apdxsection}
\renewcommand\appendix{\par
	\setcounter{savesection}{\value{section}}%
	\setcounter{section}{\value{apdxsection}}%
	\setcounter{subsection}{0}%
	\gdef\thesection{\@Alph\c@section}}
\newcommand\unappendix{\par
	\setcounter{apdxsection}{\value{section}}%
	\setcounter{section}{\value{savesection}}%
	\setcounter{subsection}{0}%
	\gdef\thesection{\@arabic\c@section}}
\makeatother

\date{December 18, 2020}

\begin{document}

\begin{frontmatter}



\title{AdaPT: Adaptable Particle Tracking for Spherical Microparticles in Lab on Chip Systems\tnoteref{t1}}
\tnotetext[t1]{© 2020. This manuscript version is made available under the CC-BY-NC-ND 4.0 license \texttt{http://creativecommons.org/licenses/by-nc-nd/4.0/}}


\author[a]{Kristina Dingel\corref{author}}
\author[b]{Rico Huhnstock}
\author[b]{Andr\'{e} Knie}
\author[b]{Arno Ehresmann}
\author[a]{Bernhard Sick}

\cortext[author] {Corresponding author.\\\textit{E-mail address:} kristina.dingel@uni-kassel.de}
\address[a]{Intelligent Embedded Systems, Department of Electrical Engineering and Computer Science, University of Kassel, Wilhelmsh{\"o}her Allee 73, 34121 Kassel, Germany}
\address[b]{Institute of Physics and Center for Interdisciplinary Nanostructure Science and Technology, University of Kassel, Heinrich-Plett-Stra{\ss}e 40, 34132 Kassel, Germany}

\begin{abstract}
Due to its rising importance in science and technology in recent years, particle tracking in videos presents itself as a tool for successfully acquiring new knowledge in the field of life sciences and physics. Accordingly, different particle tracking methods for various scenarios have been developed. In this article, we present a particle tracking application implemented in Python for, in particular, spherical magnetic particles, including superparamagnetic beads and Janus particles. 
In the following, we distinguish between two sub-steps in particle tracking, namely the localization of particles in single images and the linking of the extracted particle positions of the subsequent frames into trajectories.
We provide an intensity-based localization technique to detect particles and two linking algorithms, which apply either frame-by-frame linking or linear assignment problem solving. Beyond that, we offer helpful tools to preprocess images automatically as well as estimate parameters required for the localization algorithm by utilizing machine learning. As an extra, we have implemented a technique to estimate the current spatial orientation of Janus particles within the x-y-plane. Our framework is readily extendable and easy-to-use as we offer a graphical user interface and a command-line tool. Various output options, such as data frames and videos, ensure further analysis that can be automated.

\end{abstract}

\begin{keyword}
Particle tracking \sep Machine learning \sep Python code \sep Superparamagnetic beads \sep Janus particle \sep Magnetic particle
\end{keyword}

\end{frontmatter}

{\bf PROGRAM SUMMARY}

\begin{small}
\noindent
{\em Program Title:} AdaPT                                         \\
{\em CPC Library link to program files:} (to be added by Technical Editor) \\
{\em Developer's repository link:} \texttt{https://git.ies.uni-kassel.de/adapt/adapt} \\
{\em Code Ocean capsule:} (to be added by Technical Editor)\\
{\em Licensing provisions:} MPL-2.0                                 \\
{\em Programming language:} Python 3.6                                   \\
{\em Supplementary material:}
We provide supplementary material to increase the traceability of the provided example. It consists of an exemplary input video, the corresponding annotated video with tracked particles, a data frame including the tracking information, and a plot displaying the trajectories.                                  \\
{\em Nature of problem (approx. 50-250 words):}\\
Particle tracking in videos is an important tool for acquiring new knowledge in diverse fields. Several particle tracking methods have been developed for these diverse applications. The presented particle tracking software has been developed for the motion analysis of spherical or close to spherical magnetic particles. Up until now, no easily extensible automated particle tracking software for close to spherical microparticles and their current positioning status is available. 
\\
{\em Solution method (approx. 50-250 words):}\\
AdaPT is an extensible, easy-to-use microparticle tracking application developed explicitly for lab on chip applications but easily extensible to other applications and further functionalities. Currently implemented linking algorithms are a frame-by-frame linking approach as well as an approach solving linear assignment problems. In addition to many assistance possibilities for the user in the form of estimates of parameter values through machine learning, we offer the particular option to determine the orientation and rotation of spherical polymer particles with hemispherical metallic caps (Janus particles). The application can be used via console and graphical user interface.
\\
{\em Additional comments including Restrictions and Unusual features (approx. 50-250 words):}\\
This software requires video data with spherical or close to spherical magnetic particles. It was tested on videos containing spherical superparamagnetic and magnetic Janus particles.
Only mobile particles are detected; immobile particles are ignored by the software, reducing the amount of output data considerably.
As a unique feature, the spatial orientation within the x-y-plane of Janus particles can be determined.
The application has been tested on a variety of two-dimensional particle motion patterns.
The latest version of \textit{AdaPT} can be found here: \texttt{https://git.ies.uni-kassel.de/adapt/adapt}
\\

\end{small}

\section{Introduction} \label{sec:introduction}

Particle tracking is a crucial tool in diverse fields of fundamental science and technology. Besides applications in aerodynamics \cite{Kompenhans} and fluid dynamics \cite{Ouellette, KelleyAndOuellette}, microparticle tracking is particularly useful for the determination of microfluidic flows or in lab on chip systems with actively propelled particles, the latter often being used for the transport, separation, and detection of biological cargo \cite{Shen2017, Abdulkarim2015}.

Our focus is on tracking spherical superparamagnetic microparticles, including Janus particles, that move within tailored magnetic landscape on top of magnetically structured and topographically flat chip substrates and a dynamically varying external magnetic field \cite{Holzinger2015, Ehresmann2015}. The magnetic field landscapes of these chips may differ. For the tests of the software, we used magnetic field landscapes generated by, e.g., stripe, checkerboard, or honeycomb-like domain patterns, cf. Fig.~\ref{fig:examplesSelection}.

Since the challenge of particle tracking has been known for a long time, many tracking applications have been developed, such as the \textit{Video Spot Tracker Manual}\footnote{http://cismm.web.unc.edu/resources/software-manuals/video-spot-tracker-manual/}, the \textit{Neural Net Tracker}\footnote{https://aitracker.net/} \cite{Newby2018}, or the \textit{ParticleTracker}\footnote{http://mosaic.mpi-cbg.de/?q=downloads/imageJ} \cite{Xiao2016}, to name a few. In our case, however, the objectives are very specific:
For further, much more significant, analysis and evaluation, it is particularly required to track as many particles as possible. Not only the particle positions and, therefore, tracks are of relevance, but other features such as the current spatial orientation of a particle are needed to fully understand the particle movement behavior.

Therefore, we have developed a framework for particle tracking, namely \textit{AdaPT}, which we will present in the following.

In general, we consider particle tracking in two separate steps: the localization of the particle positions in the respective frames of a video and the linking of the extracted particle positions into probable trajectories. We offer an intensity-based localization method and two linking algorithms that utilize either frame-by-frame linking methods or linear assignment problem solving.

Our goal is to create a tracking technique that is easy-to-use and extensible for further requirements, for example, the determination of the full 3D position, trajectory, and velocity at a later stage. Besides, we have reduced the amount of required user input by utilizing machine learning techniques, thus enabling automated tracking. As an extra, we offer to determine the orientation of hemispherical metallic caps of polymeric Janus particles. For further analysis, it is particularly important that the outputted information can be processed further in an automated manner, which is why this is another main focus of this application.

In Sec.~\ref{sec:scenario}, we provide details of the requirements and consequent restrictions regarding the developed tool, which are presented by means of our own application scenario.
Afterward, the implementation of the application is described in more detail in Sec.~\ref{sec:impl}. Sec.~\ref{sec:execution} displays three feasible ways to run the application, whereas Sec.~\ref{sec:example} provides an exemplary procedure to use the framework.

\section{Requirements and Restrictions} \label{sec:scenario}

The framework was designed and optimized to determine 2D trajectories of spherical magnetic particles placed and propelled on top of a magnetically patterned and topographically flat chip substrate. The current prototype substrates consist of an Exchange Bias (EB) thin film system with magnetic domains of designed shapes and magnetization orientations, which are patterned into the system via ion bombardment induced magnetic patterning \cite{Mougin2001, Ehresmann2004, Gaul2016}. 

Transport experiments were conducted by placing an aqueous solution of magnetic particles on top of the patterned EB substrates by superimposing an external magnetic field pulse sequence. In some cases, an additional polymeric layer with a thickness of about 700 $nm$ was placed between substrate and particles to avoid sticking of the particles to the substrate surface. As a direct consequence of this transport concept, which aims to be applied in future lab on a chip devices \cite{Holzinger2015, Ehresmann2015}, a step-wise motion of the particles is observed.

\begin{figure}[!htbp]
	\centering
	\subfloat[Large particles that appear hollow.]{\includegraphics[width=6cm]{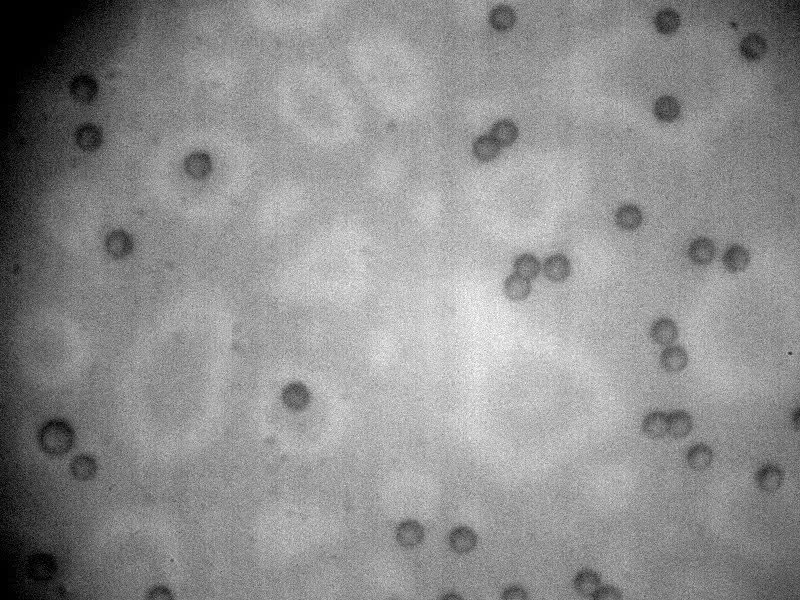}}\qquad
	\subfloat[Unsorted small particles.]{\includegraphics[width=6cm]{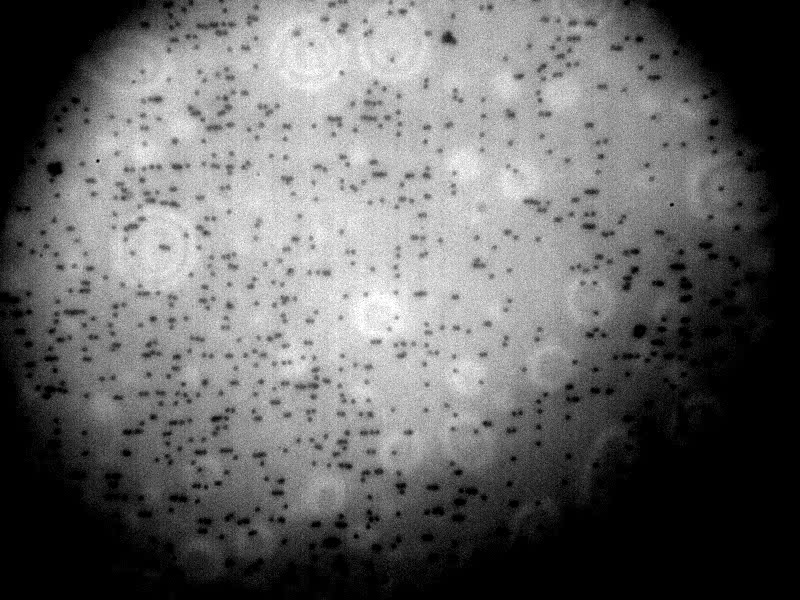}}\qquad
	\subfloat[Small particles sorted within a parallel-stripe domain pattern.]{\includegraphics[width=6cm]{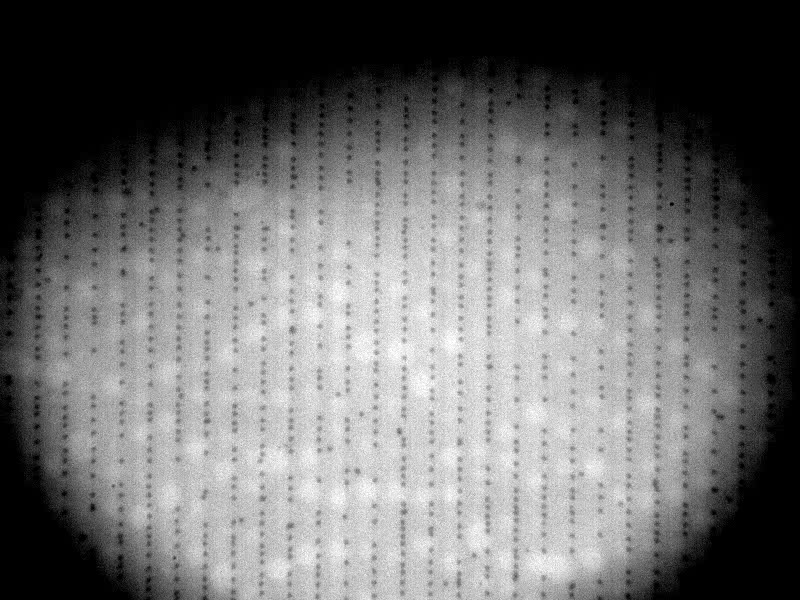}}\qquad
	\subfloat[Particles within a honeycomb pattern.]{\includegraphics[width=6cm]{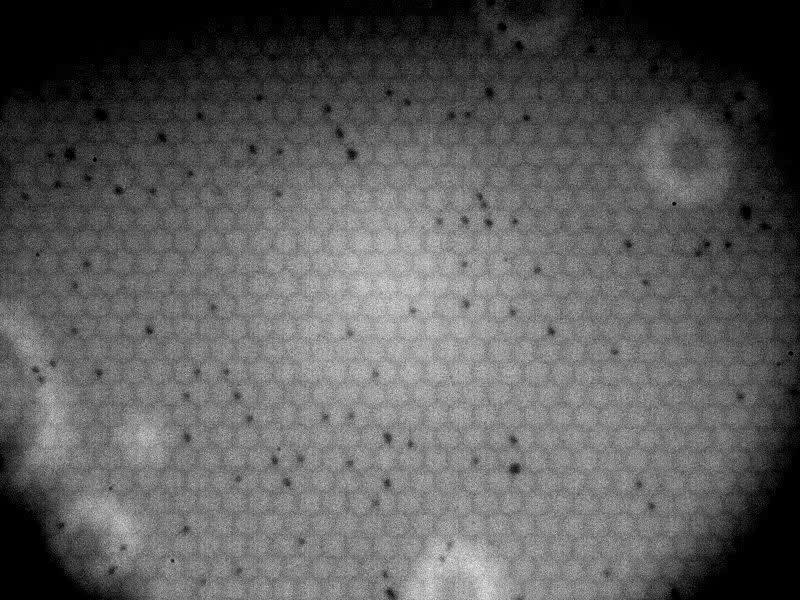}}\qquad
	\subfloat[Particles groups building a chess pattern.]{\includegraphics[width=6cm]{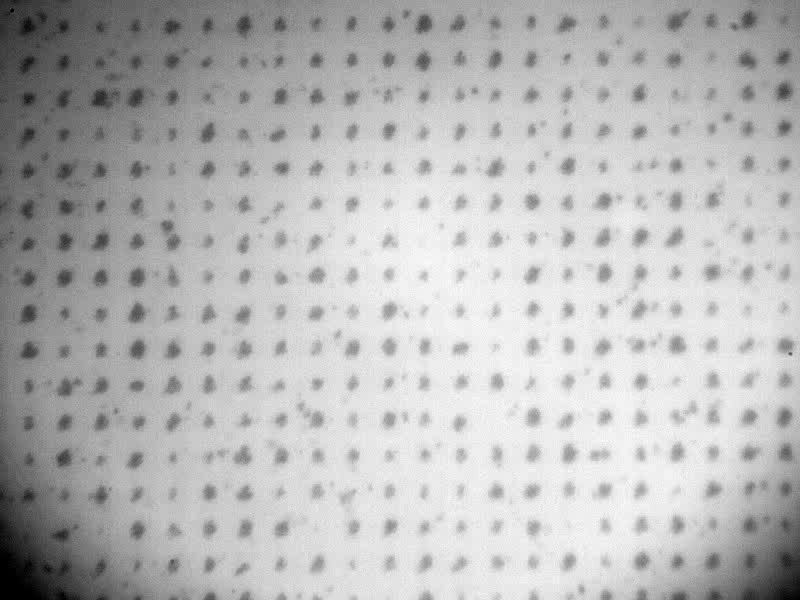}}\qquad
	\subfloat[Janus particles.]{\includegraphics[width=6cm]{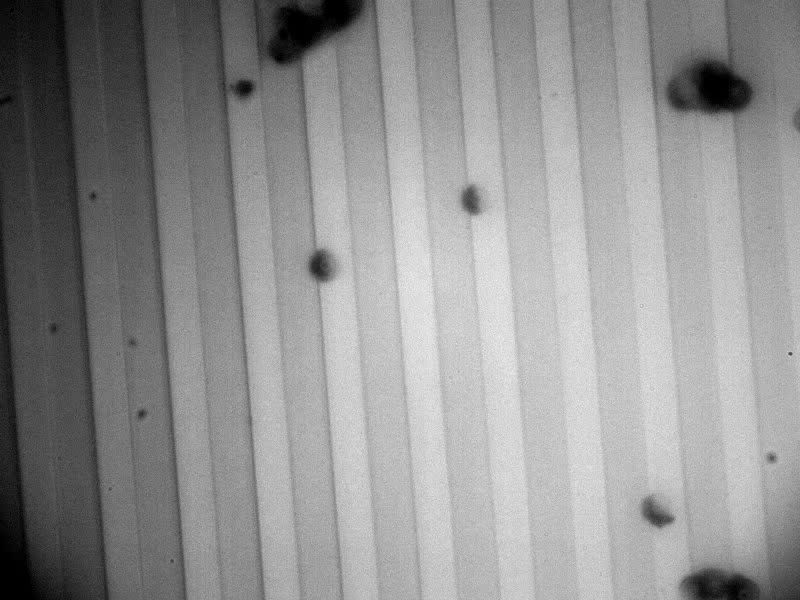}}
	\caption{Sample selection of feasible particle tracking input images.}
	\label{fig:examplesSelection}
\end{figure} 

Since an accurate description of the particles trajectories is crucial for understanding the underlying fundamental physical processes, this application's essential requirement is to precisely track all visible and mobile particles during a recorded transport video. 
	In our case, only mobile particles are of interest for further analysis since we want to investigate particle movement patterns for application in future lab on chip systems.
This then allows for systematically mapping the influence of varying experimental parameters on the particle motion, i.e., the particle transport sensitivity against external stimuli can be determined.

We used various domain patterns and particle classes to test the application. A selection is provided in Fig.~\ref{fig:examplesSelection}. The domain patterns range from a simple parallel-stripe configuration for one-dimensional translation of the particles, cf. Fig.~\ref{fig:examplesSelection} (c), to more complex checkerboard, cf. Fig.~\ref{fig:examplesSelection} (e), and honeycomb-like, cf. Fig.~\ref{fig:examplesSelection} (d), structures that allow for two-dimensional transport. Depending on the chosen domain pattern, the application can track single particles, cf. Fig~\ref{fig:examplesSelection} (a)-(d), and also bunches of multiple particles traveling together across the substrate, cf. Fig~\ref{fig:examplesSelection} (e). In these cases, we used superparamagnetic core-shell particles with diameters ranging between 1 $\mu$$m$ and 6 $\mu$$m$.

Moreover, the application allows for monitoring the orientation and rotational motion of magnetic Janus particles in the x-y-plane, cf. Fig~\ref{fig:examplesSelection}~(f). These Janus particles are composed of non-magnetic silica beads and a magnetic hemispherical cap on top, which was fabricated by depositing an EB thin film system onto a monolayer of silica beads. Due to the magnetization orientation within the magnetic cap, the Janus particles exhibit a combined translational and rotational behavior during a transport experiment. To 
describe the observed spatial orientation of the cap, we offer a tracking feature for Janus particles.

Although this tool was designed for the above-described scenario, tracking should be possible for other cases that meet the following characteristics:

\begin{itemize}
	\item 
	The implemented algorithm should be able to track all kinds of Gaussian-like particles. However, in case one does not use almost spherical particles, the option to "fill" particles should be switched off (cf. Sec.~\ref{impl:preprocessing}), as it only works for circular shapes.
	\item We recommend using particles with approximately the same size within a video
	, as the particle size is required as a parameter for the localization algorithm (cf. Sec.~\ref{impl:localization}). It assures that only suitable particle candidates are located and a sufficient subpixel accuracy is ensured.
	\item The application automatically disregards immobile particles 
	as in the application case described above, only mobile particles are of interest. However, with some modification of the code, the background subtraction, which eliminates the detection of immobile particles, can be discarded.
	\item Currently, dark particles are detected on a light background. With some modifications of the source code, light particles on a dark background can be detected as well.
\end{itemize}

\section{Implementation Details} \label{sec:impl}

In order to comprehend the functionality of the application, the applied algorithms and strategies are described in this section. Hereafter, we provide an overview of the application pipeline, which is described in more detail below.

\begin{figure}[h]
	\centering
	\includegraphics[width=\textwidth]{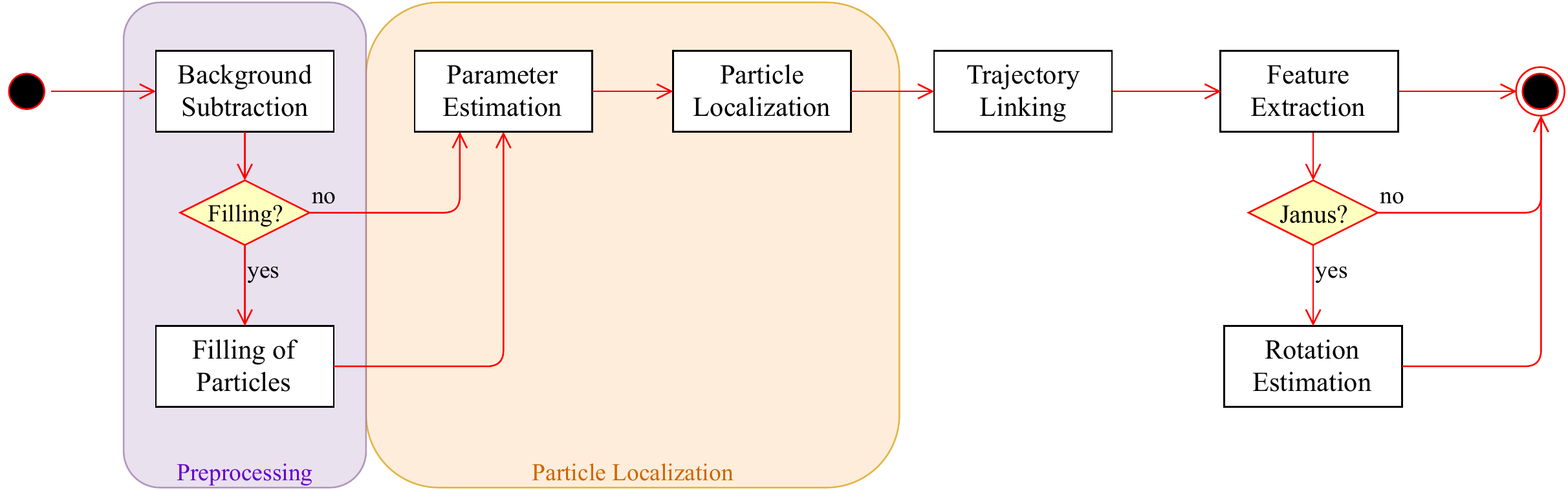}
	\caption{Schematic application pipeline of AdaPT.}
	\label{fig:pipeline}
\end{figure}

The pipeline of \textit{AdaPT} consists of four stages. In the initial stage, the input video is preprocessed such that particles can be more easily located in the subsequent step. As stated before, a requirement of this application is that only the characteristics of moving particles have to be analyzed. That is why a background subtraction (subtraction of the average image of a video from the current frame) is applied within the preprocessing stage. As the localization algorithm requires the particle center to be the most intense, a "filling" of particles with appearing hollow kernel or 
blurry images is advantageous, cf.~Fig.~\ref{fig:examplesSelection} (a) and (f), and the used localization algorithm requires the center of the particle to be the most intense. Particles are then localized using the algorithm proposed by Crocker and Grier~\cite{Crocker1996}. To execute this algorithm, user-defined parameters are required, which can be automatically estimated beforehand. The determined particle positions of successive frames are then used to link them into probable trajectories. Two linking algorithms can be used: a frame-by-frame linking approach by Crocker and Grier~\cite{Crocker1996} as well as a linear assignment problem solver by Jaqaman et al.~\cite{Jaqaman2008}. In the last stage, optional features such as the current orientation of Janus particles can be determined and attached to the output data containing, e.g., the trajectories.
Fig.~\ref{fig:pipeline} shows a schematic representation of the pipeline described above.

\subsection{Intelligent Preprocessing} \label{impl:preprocessing}

As stated before, for each image of a video, background subtraction is applied. Therefore, a pixel-by-pixel intensity average of all video frames is determined and subtracted from the current frame. This step is necessary due to the noisy background of the images, cf. Fig.~\ref{fig:examplesSelection}. This step removes immobile particles from the images, thus reducing the output data only to information corresponding to mobile particles. 

As a further option, a preprocessing of the frames where particles appear hollow or blurry, cf.~Fig.~\ref{fig:examplesSelection}~(a) and (f), is offered such that they are better recognizable by the localization algorithm in the subsequent step. Here, the "filling" of the particles may be of advantage. In case that filling is required, we use \textit{Hough Circle Detection} \cite{Bradski2008} to automatically detect the circles and fill them with the brightest color, which in our case is white.

In addition to the possibility of selecting the required preprocessing steps \textit{background subtraction only} and \textit{additional filling of the particles} manually, we provide an 
\textit{ensemble random forest classifier} \cite{BreimanLeo2001}, which can predict if "filling" of the particles is required according to characteristics extracted from the given videos.  
Appendix \ref{ap:features} sums up the utilized characteristics.

In our case, further preprocessing steps are carried out within the subsequent localization algorithm.

\subsection{Particle Localization} \label{impl:localization}

This section focuses on identifying probable particle candidates within the preprocessed images. Mainly, the localization algorithm is demonstrated. Furthermore, the implemented parameter estimates regarding the algorithm are introduced. 

\subsubsection{Algorithm}

Particles are localized using a greedy algorithm by Crocker and Grier~\cite{Crocker1996}. 
In the first stage of this algorithm, the input data is preprocessed by a spatial bandpass filter with small-scale variation such that camera noise is reduced. Additionally, a bandpass filter with large-scale variation is required to correct uneven lighting. 
After preprocessing, local maxima are determined by comparing the brightness of the pixels. These local maxima form probable particle candidates. In case two maxima are too close to each other, only the brightest particle will be chosen as a particle candidate.
To estimate the exact center of the particles with subpixel precision, the intensities of the pixels close to the particle candidates are averaged and weighted by brightness. 

Although this algorithm is considered to run in an automated fashion, it cannot do that without user input. In total, two parameters, namely the \textit{particle size} and the \textit{minimum mass} of the particle, are required to achieve satisfying localization results. In this case, the particle size is the (estimated) size of the particles in pixels. The minimum mass is the total integrated brightness within the pixels representing the particle area.
The proposed algorithm was implemented using the library \textit{Trackpy} \cite{Allan2018}, which implements and provides, e.g., the described localization method by Crocker and Grier \cite{Crocker1996}.

Since it is cumbersome to estimate these parameters manually, a machine learning based method was developed for an automated parameter estimate.

\subsubsection{Automated Parameter Estimate} \label{impl:parameters}

To estimate the average size and a minimum mass of the particles, characteristics from the videos are required again to train the according regression models. Separate logistic regression models were trained for both required parameters by using the characteristics in Appendix~\ref{ap:features}.

\subsection{Trajectory Linking} \label{impl:linking}

Two algorithms are implemented to link extracted particle positions into trajectories. The first is a frame-by-frame algorithm that can link particles into trajectories without much computational effort. The second is a linear assignment problem solver, which can additionally detect gaps, merges, and splits between trajectories.

\subsubsection{Frame-by-Frame Linking}

For their linking algorithm, Crocker and Grier \cite{Crocker1996} assume that the particles are undergoing Brownian motion.
This leads to the assumption that the particle position in the next frame will most likely be next to the position in the
current frame. Particles are connected into trajectories by minimizing the total length of the links.
This algorithm allows particles to leave or enter the frame. Additionally, particles
can be tracked if they temporarily vanish and reappear nearby. This may occur, e.g., when particles are too dark to be detected in single frames or, for the present application, when they stick to the surface of the substrate for a series of magnetic field pulses and then move again.
Again, the library \textit{Trackpy} \cite{Allan2018} is utilized to implement this linking approach within the application.

\subsubsection{Linear Assignment Problem Solver}

Jaqaman et al. \cite{Jaqaman2008} provide a robust and computationally feasible approximation of multiple hypothesis tracking \cite{Reid1978} by using a two-step approach to link particle positions into probable trajectories. Given a set of detected particles, this algorithm links particles frame-by-frame in the first run. During the second run, the trajectory segments are revisited to close gaps, capture particle merging and splitting. The first step is greedy, though optimal in space, whereas the second step is performed using temporally global optimization. Both steps are formulated as a linear assignment problem, where the aim is to find the combination of assignments with the minimal sum of costs. This algorithm is independent of the dimensionality, particle motion, and the physical nature of the particle.

The framework \textit{lap\_tracker} \cite{LAPTracker} provides a Python implementation of this algorithm, which is used within \textit{AdaPT}. Due to obstacles encountered with the operating system Windows in conjunction with \textit{lap\_tracker}, we provide a version of \textit{AdaPT} without this algorithm. For details, we refer to the \texttt{README} file.

\subsection{Optional Feature: Janus Particle Orientation Estimate} \label{impl:rotation}

\begin{figure}[!htbp]
	\centering
	\subfloat[Theoretical appearance of a Janus particle and used intensity profiles including angles $\alpha$ for rotation estimation in the x-y-plane.]{\includegraphics[width=5.5cm]{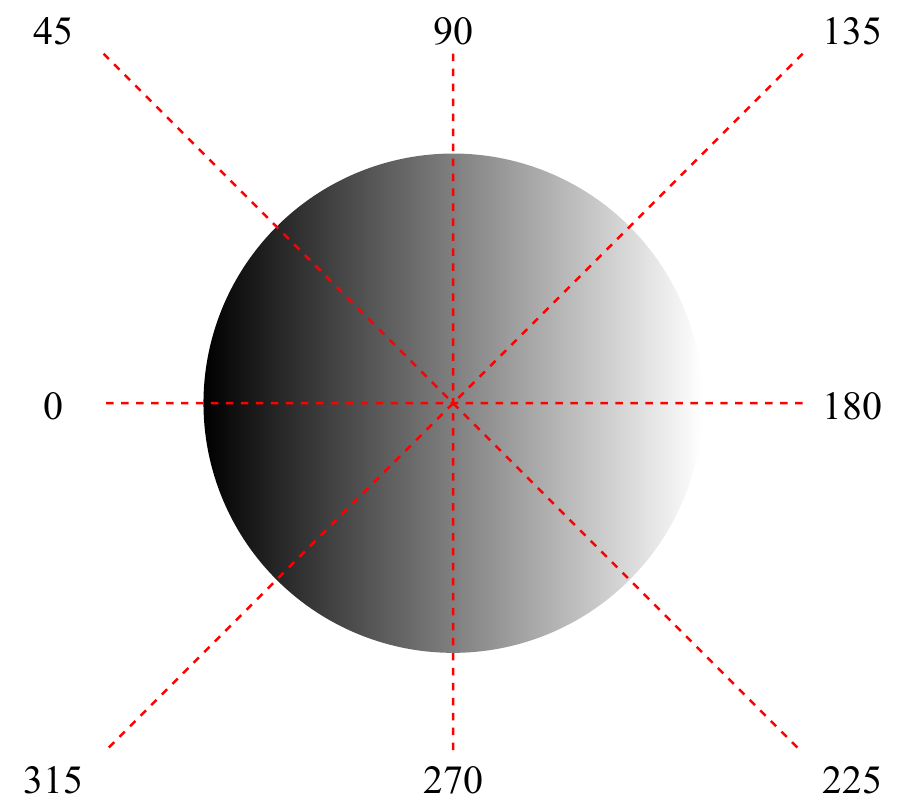}}\qquad
	\subfloat[Real Janus particle after background subtraction and cropping.]{\includegraphics[width=5.5cm]{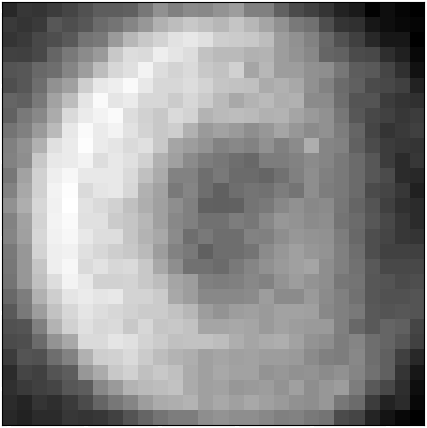}}\qquad
	\qquad
	\subfloat[Diagonal intensity profile from the left upper corner to the right lower corner.]{\includegraphics[width=5.5cm]{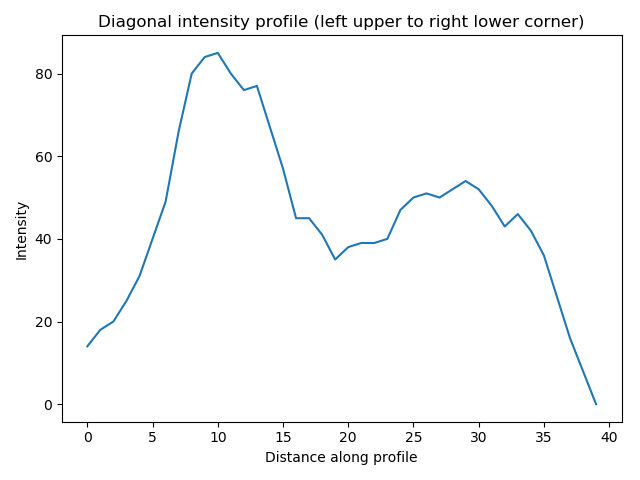}}\qquad
	\subfloat[Diagonal intensity profile from the right upper corner to the left lower corner.]{\includegraphics[width=5.5cm]{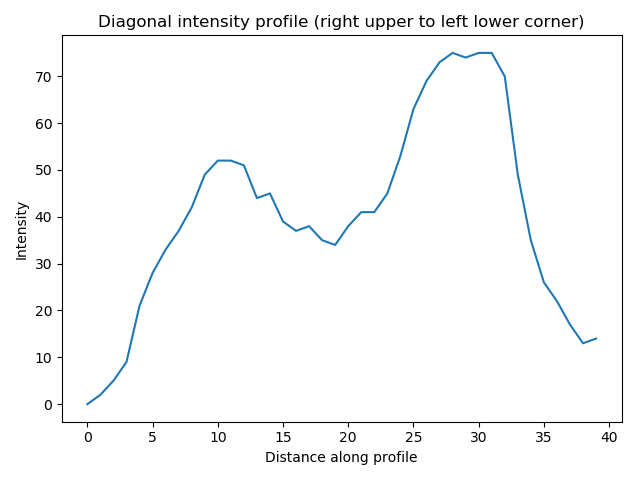}}\qquad
	\subfloat[Horizontal intensity profile from left to right.]{\includegraphics[width=5.5cm]{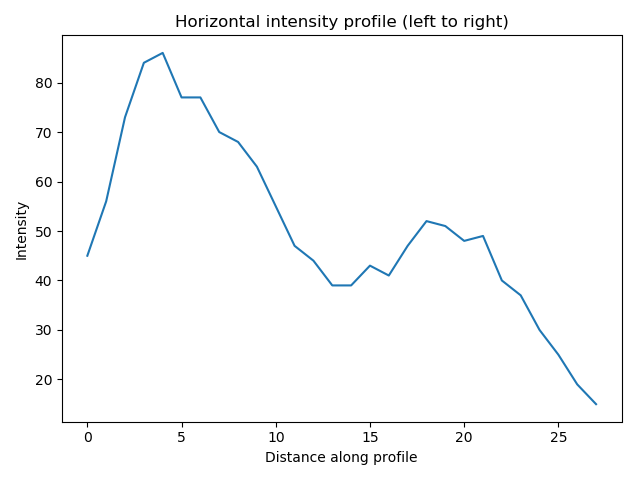}}\qquad
	\subfloat[Vertical intensity profile from top to bottom.]{\includegraphics[width=5.5cm]{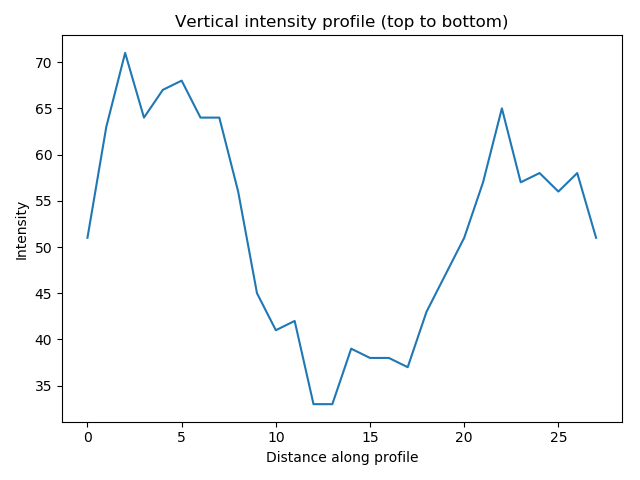}}
	\caption{Exemplary representation of the functionality of the orientation estimate.}
	\label{fig:janus}
\end{figure} 

As stated by Horn \cite{Horn1977}, intensity profiles of an image can be used to estimate several characteristics, especially the shape of objects. We use this fact and utilize intensity profiles to estimate Janus particles' current orientation in the x-y-plane, cf. Fig.~\ref{fig:janus} (a). Intensity profiles are well suited for this kind of task because Janus particles in a sense defined above are characterized by a dark and a bright fraction. Notably, the orientation of the metallic magnetic cap (i.e., the opaque side of the Janus particle) is of interest. In the following, the images after background subtraction are used, cf. Fig.~\ref{fig:janus}~(b), which is why, in the following, the bright pixels are of interest. For this option, of course, the preprocessional filling of the particle images must not be carried out. To determine the orientation of the hemispherical cap, we have to look for maxima in the intensity profiles, i.e., the bright pixels are of interest after background subtraction, as shown in Fig.~\ref{fig:janus} (b).

The prerequisite to determine a Janus particle's orientation is that it has been localized before because in the next step, an image section containing only the particle is required. Using this cropped image, cf. Fig.~\ref{fig:janus} (b), four (or optionally more) intensity sections are determined within the image as shown in Fig.~\ref{fig:janus} (a). The rotation degrees shown next to the profile lines are required to calculate the orientation, which is entered into the resulting output data frame. To compute the orientation, maxima of the determined intensity sections are extracted. Here, one has to distinguish between two cases: either one or more maxima are detected.

If only one maximum is detected, the result of the returned value depends on the position of the maximum relative to the intensity profile. Suppose the detected maximum is close to the center of the profile. In that case, it most likely does not provide sufficient information regarding the orientation of the particle within the x-y-plane and is therefore omitted. If the detected maximum is on one side of the intensity section, the orientation angle $\alpha$ within the x-y-plane, as defined in Fig 3 (a), next to the detected maximum is stored.

In the case that two or more maxima are detected, the angle $\alpha$ to be stored depends on the ratio between the first ($\mathit{max_{1}}$) and the last detected maximum ($\mathit{max_{-1}}$) in terms of the position on the intensity profile, where $\mathit{min\_diff}$ represents the minimum required spatial distance between $\mathit{max_{1}}$ and $\mathit{max_{-1}}$:

\begin{enumerate}
	\item If $\mathit{max_{1} > max_{-1}}$ and $\mathit{|max_{1} - max_{-1}| > min\_diff}$, $\alpha$ next to $\mathit{max_{1}}$ is stored.
	\item If $\mathit{max_{1} < max_{-1}}$ and $\mathit{|max_{1} - max_{-1}| > min\_diff}$, $\alpha$ next to $\mathit{max_{-1}}$ is stored.
	\item If $\mathit{|max_{1} - max_{-1}| < min\_diff}$, the intensity profile leads to no significant information about the current orientation of the particle and is therefore discarded. In this case, the angles $\alpha$ next to the intensity profile are not required for the computation.
\end{enumerate}

Using the returned angles $\alpha$ from all intensity profiles, the resulting orientation is computed by averaging all obtained angles such that exactly one angle emerges as output. The returned average angle represents the cap orientation in the plane parallel to the substrate. An exemplary computation of the Janus particle's orientation in Fig.~\ref{fig:janus} (b) is shown in Sec.~\ref{ex:rotation}.

\subsection{Additional Implementation Details}

\emph{Parallelization:}
To enable faster processing, the localization step has been parallelized. By default, the process is parallelized depending on the available number of cores. 

\emph{Reproducability:}
It is often desired to repeat the tracking process with the same chosen values. Therefore, we provide an option to convert the parameters set in the graphical user interface (GUI) into a command for the command prompt such that it is possible to repeat the analysis.

\emph{Extensible:}
We follow an object-oriented approach that allows for a simple implementation of new algorithms or further options.

\section{Execution} \label{sec:execution}

In total, there are three possibilities to launch the application implemented in Python. These are described below and the advantages of each technique are displayed. Note that the requirements in the provided \texttt{README} file have to be fulfilled to run the application.

\subsection{Execution using a console}

One way to start \textit{AdaPT} is by changing the provided \texttt{configuration.ini} file beforehand and running \texttt{python3 -m main.main} within the root folder of the project. This way, the provided parameters are stored after the tracking process and the analysis can be repeated. Parameters within the configuration file can be set directly through command-line parameters as well. If at least the parameter \texttt{--input} is provided when running \texttt{python3 -m main.main}, the \texttt{configuration.ini} file will be ignored. With these execution methods, \textit{AdaPT} can be immediately executed on, e.g., a server.

Table \ref{tab:params} lists all available parameters that can be set by using either the configuration file or the command line parameters.

\begin{table}
	\begin{tabularx}{\textwidth}{p{4cm} p{3.5cm} X}
		\toprule
		\multicolumn{2}{c}{\textbf{Parameter}} & \textbf{Description} \\ 
		\textit{Console} & \textit{Configuration File} & \textbf{} \\
		\midrule
		\texttt{--input} & \texttt{Path} & Path to the video file. \\
		\texttt{--automated\_\newline preprocessing} & \texttt{Automated-\newline preprocessing} & Specifies, whether the preprocessing process to apply should be chosen manually. \\
		\texttt{--fill\_in\_particles} & \texttt{Fill-in} & If manual preprocessing is chosen, this assures that particles will be filled. \\
		\texttt{--janus} & \texttt{Rotation} & Computes the orientation, if Janus particles are contained in a video. \\
		\texttt{--create\_video} & \texttt{Video} & Specifies, whether a video should be created afterward. \\
		\texttt{--estimate\_params} & \texttt{Estimate-params} & Parameters \textit{particle size} and \textit{minimum mass} (see definitions above) are estimated if checked.\\
		\texttt{--particle\_size} & \texttt{Particle-size} & Particle size parameter for the localization algorithm.\\
		\texttt{--min\_mass} & \texttt{Min-mass} & Minimum mass parameter for the localization algorithm. \\
		\texttt{--frames} & \texttt{Number-of-frames} & Number of frames to be processed. If $0$ is provided, all frames will be processed.\\
		\texttt{--lap} & \texttt{LAP} & If checked, the linking algorithm by Jaqaman et al. \cite{Jaqaman2008} is used. Otherwise, the algorithm by Crocker and Grier is utilized \cite {Crocker1996}. \\
		\bottomrule
	\end{tabularx}
	\centering
	\caption{Parameters that can be set through either console or a configuration file for which a template is provided.}
	\label{tab:params}
\end{table}

\subsection{Execution using a GUI}

\begin{figure}[!htbp]
	\centering
	\includegraphics[width=\textwidth]{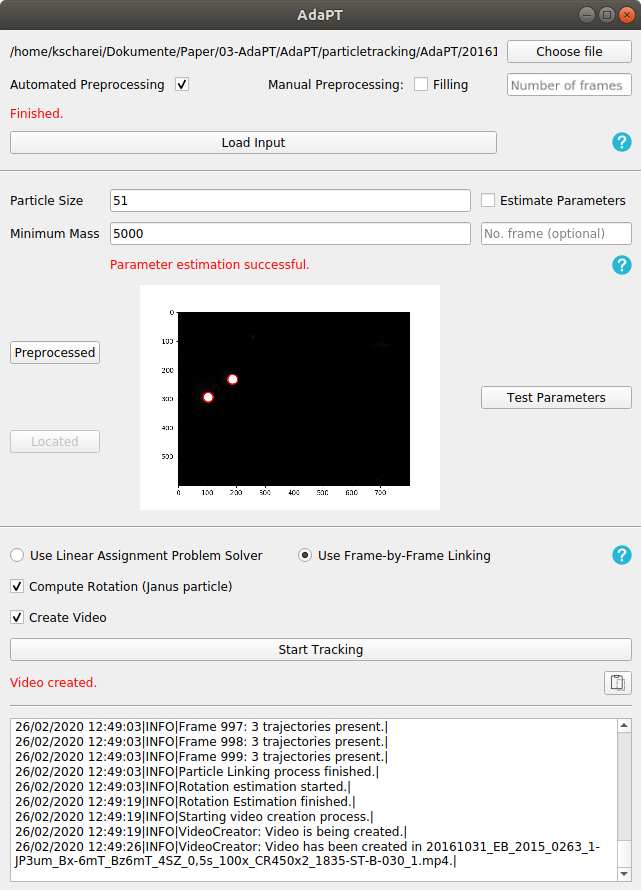}
	\caption{Resulting GUI of \textit{AdaPT} after performing the example.}
	\label{fig:UI}
\end{figure}

Furthermore, a GUI can be used to track particles. The layout of the GUI is shown in Fig~\ref{fig:UI}. The top section of the GUI deals with the import of the video. After choosing a video file, the preprocessing of the video's images can either be manually selected or automatically controlled. If desired, the number of frames to be analyzed can be specified before loading the input. Once the video has been loaded, parameters regarding the localization algorithm can be set in the next section. These may either be estimated or entered manually in the corresponding fields. Once the \texttt{Test Parameters} button has been pushed, a frame including the localized particles is displayed. The preprocessed image may be viewed as well. If desired, the frame to be tested can be chosen manually by entering its number into the \texttt{No.~frame} field. By default, the first frame of the video is used. Once satisfactory parameters have been selected, the particle tracking can start in the next section by using either frame-by-frame linking or linear assignment problem solving. Optionally, the orientation of Janus particles can be estimated or a video, including the tracking result, can be created. The button beneath the \texttt{Start Tracking} button copies the corresponding console command to the clipboard such that the analysis can be reproduced afterward. In the lower text field, the logging outputs are displayed in order to track the progress of the application. The question marks next to the individual sections provide helpful tooltips on how to use the single GUI sections.

The GUI was developed for the purpose of testing parameters in a simple and, therefore, straightforward way by viewing the output immediately. Accordingly, the GUI can be used to determine correct parameters, e.g., \textit{particle size} and \textit{minimum mass}, regarding new input data.

More detailed descriptions on how to use \textit{AdaPT} can be found in the \texttt{README} file. In the next section, we provide an exemplary procedure on how to use the application.

\section{Exemplary Procedure} \label{sec:example}

To illustrate how the program operates, this section exemplary describes the application procedure for Janus particles. Figure~\ref{fig:UI} displays the resulting GUI after applying the beneath described procedure.

\subsection{Input}

To start the application, a video to be analyzed is expected. The application was tested with videos in \texttt{AVI} format, but other formats such as \texttt{MP4} can be analyzed as well. Figure~\ref{fig:examplesSelection} displays a selection of different particle videos that can be analyzed using \textit{AdaPT}. 

In the following, the example will be referring to the sample video containing Janus particles in Fig.~\ref{fig:examplesSelection} (f). (Sample video \texttt{example.avi} is provided in the supplementary material). 

\subsection{Preprocessing}

After selecting the video to be analyzed, corresponding preprocessing steps need to be determined. Therefore, either manual or automated preprocessing may be considered. As the classification model, cf.~Sec.~\ref{impl:preprocessing}, is trained on these types of data, cf.~Fig.~\ref{fig:examplesSelection}, we choose automated preprocessing. The classifier's output is class $1$, which stands for \textit{preprocessing using background subtraction and particle filling}. In this example, we have limited the analysis to a thousand frames.

\subsection{Parameter Estimate}

The next step involves estimating the parameters required for the localization algorithm. Either the parameters can be automatically estimated or typed in manually. Furthermore, the preprocessed image or the image with localized particles can be visualized. If new data is provided and the parameter estimate does not work, this way, parameters can be tested manually in a straightforward way. For example, especially this visualization tool within the GUI was utilized to determine the correct labels for the training of the regression models that estimate the input parameters of the localization algorithm, cf. Appendix \ref{ap:regression}.

\subsection{Localization and Linking}

After satisfactory results regarding the parameters have been achieved, the actual tracking process can be initialized. Therefore, it is essential to specify which kind of linking algorithm should be used. In our example, we choose the frame-by-frame linking approach. In addition, we want to compute the orientation of the Janus particles in each frame and create a video of the tracking result afterward. These two features are explained in more detail in the following sections.

\subsection{Rotation Estimate} \label{ex:rotation}

After the basic principles of determining the current orientation of Janus particles in Sec.~\ref{impl:rotation} have been clarified, the orientation for the Janus particle in Fig.~\ref{fig:janus} (b) is hereafter calculated as an example using $\mathit{min\_diff=14}$, with $\mathit{14}$ being half the width of the cropped image. In general, $\mathit{min\_diff}$ is always set to half of the cropped image's width. The cropped image of the Janus particle is taken from the exemplary video. 

In the intensity profile in (c), which is applied from the left upper corner to the right lower corner, two local maxima are detected at position $9$ ($\mathit{max_{1}}$) and position $28$ ($\mathit{max_{-1}}$). The values at these points are compared to each other. Since the value of $\mathit{max_{1} > max_{-1}}$ and $\mathit{|max_{1} - max_{-1}| > min\_diff}$, the angle $\alpha$ next to $\mathit{max_{1}}$, namely $45^{\circ}$, is stored, cf. Fig~\ref{fig:janus} (a). By following this calculation pattern, the returned $\alpha$ of (d) is $315^{\circ}$, the returned $\alpha$ of (e) is $0^{\circ}$, and after analyzing (f) there is no significant information and is therefore not included in the calculation of the rotation. The average angle of the obtained angles $45^{\circ}$, $315^{\circ}$, and $0^{\circ}$ equals to the output $0^{\circ}$. 

\subsection{Output}

There are two types of outputs that can result from the application. The tracking data is always stored in a data frame using \texttt{.h5} format and can be used for further (automated) processing. As a visual aid, it is optionally possible to create a video with the tracked particles. This section sums up how the output looks like in a real case.

\subsubsection{Data Frame}

\begin{sidewaysfigure}[!htbp]
	\centering
	\includegraphics[width=\textwidth]{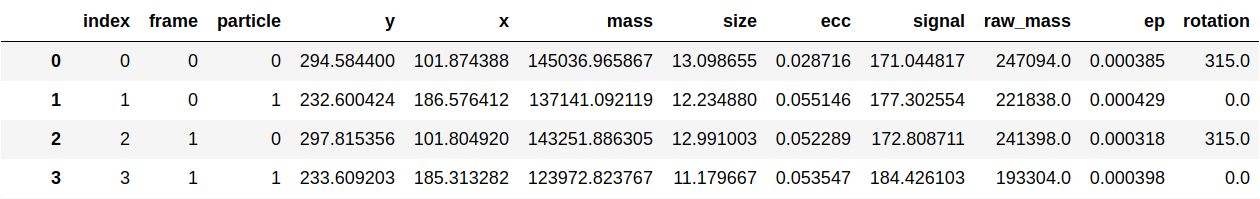}
	\caption{Format of the resulting output data frame.}
	\label{fig:h5result} 
\end{sidewaysfigure}

Figure~\ref{fig:h5result} shows an exemplary data frame after trajectory linking. Here, \texttt{frame} displays the frame number and \texttt{particle} denotes all detected particles in this frame. If particles belong to one trajectory, the same number is assigned in column \texttt{particle}. For example, the particle in line 0 is linked to the particle in line 2 in the data frame. Columns \texttt{x} to \texttt{ep} include information as displayed in Tab.~\ref{tab:trackpy}. Further information regarding the other columns can be found in the documentation of \textit{TrackPy} \cite{Allan2018}.

\begin{table}
	\begin{tabularx}{.8\textwidth}{l X}
		\toprule
		\textbf{Value}      & \textbf{Description} \\ \midrule
		$x$ & $x$ coordinate of the particle \\
		$y$ & $y$ coordinate of the particle \\
		mass & total integrated brightness of the blob \\
		size & radius of gyration of the blob's Gaussian-like profile \\
		ecc & eccentricity (0 is circular) \\
		raw mass & total integrated brightness in raw image \\
		signal & signal strength \cite{UncertaintyTrackpy}\\
		ep & estimate of uncertainty \cite{UncertaintyTrackpy} \\
		\bottomrule
	\end{tabularx}
	\centering
	\caption{Description of the returned features after localization in \textit{Trackpy} \cite{Allan2018}.}
	\label{tab:trackpy}
\end{table}

The column \texttt{rotation} is added if the orientation of Janus particles in the x-y-plane is computed. 
The resulting data frame for this example can be found in \texttt{example\_data\_frame.h5}.

\subsubsection{Video}

\begin{figure}[!htbp]
	\centering
	\includegraphics[width=.8\textwidth]{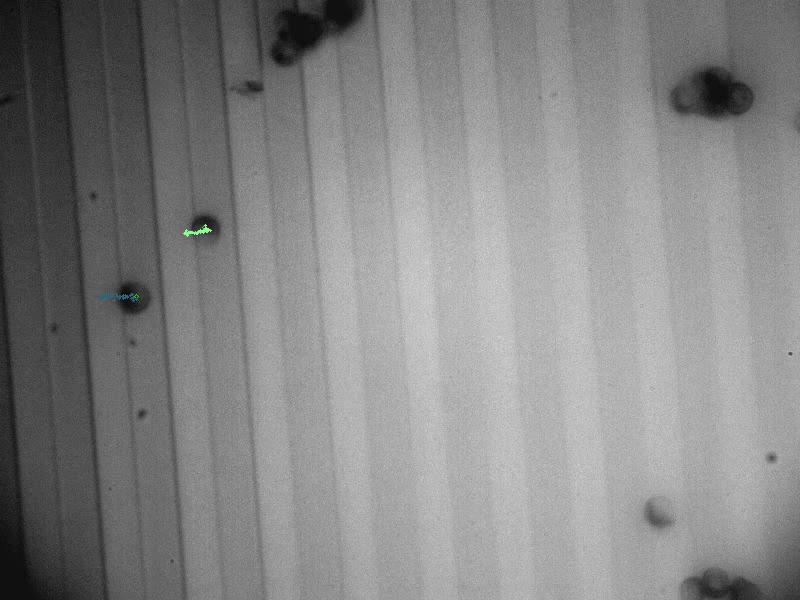}
	\caption{Snapshot of the annotated video containing two moving Janus particles.}
	\label{fig:example_annotated} 
\end{figure}

Next to a data frame, a video can be generated if required. Usually, up to a thousand frames, starting with the first frame, are annotated and outputted as a video. This is due to computation reasons. The last hundred localization points of each particle are drawn as dots into the current frame by using different colors for each particle.
The resulting video for this example can be found in \texttt{example\_annotated.mp4}, cf.~Fig.~\ref{fig:example_annotated}. Another visual representation of trajectories can be seen in \texttt{example\_trajectories.png}.

\section{Conclusion}

In this article, we provided a particle tracking framework for, initially, spherical magnetic particles, which can be run by using a GUI or a console. Besides one localization and two linking algorithms, we provided tools to automatically estimate parameters regarding the algorithms using machine learning techniques and extract further features such as the spatial orientation of Janus particles within the x-y-plane. We designed the application to be extended in the future with additional particle tracking algorithms and further options.

Although there are numerous options to extend \textit{AdaPT}, in the future, our focus is on developing new algorithms for particle tracking based on deep learning strategies.
This involves not only 2D tracking but 3D tracking as well. For further advanced analysis of the trajectories, it will undoubtedly be necessary to perform 3D tracking instead, which applies in particular to the orientation computation of Janus particles in the original 3D space. Therefore, it is inevitable to first look upon extraction methods for the missing z-coordinate by using, e.g., image gradient filters. Afterward, this new information will be used to extend the tracking algorithms mentioned in this article, or even to create new, more general, algorithms such as neural networks.

\section*{Acknowledgment}

Parts of this research were supported by the Hesse State initiative LOEWE 3 (HA-Project-No. 576/17-58). We gratefully acknowledge the assistance and support of the Joint Laboratory AIM-ED.

\appendix

\section{Features for Classification and Regression} \label{ap:features}

In order to detect and select distinguishable characteristics for the choice of a correct preprocessing procedure as well as parameter estimates, the provided videos in Fig.~\ref{fig:examplesSelection} have to be further studied. In this case, precisely those characteristics must be detected that can assign the correct required labels to the provided videos. The question of which preprocessing procedure to choose is a binary classification problem, where the question arises: Is "filling" of the particles necessary? In the case of parameter estimation, namely \textit{particle size} and \textit{minimum mass} in a sense defined above, natural numbers are possible outputs, which is why regression is appropriate at this point.

Note that the upcoming characteristics required for the training of the classifier and regression models have been extracted from the corresponding videos in Fig.~\ref{fig:examplesSelection}. If the videos to be analyzed differ from these, the models' returned predictions can be arbitrary. In this case, the model must be retrained by using the new data and possibly new features. More information regarding the retraining of the models can be found in the provided \texttt{README} file.

\subsection{Classification} \label{ap:classification}

In the case of preprocessing sequence classification, either class $0$ \textit{(Apply only background subtraction)} or class $1$ \textit{(Apply background subtraction and particle filling)} should be returned after providing a video frame.

Consider again all application cases in Fig.~\ref{fig:examplesSelection}. The first and most straight forward feature to distinguish the two classes is the size of the particles. Bigger particle images tend to appear hollow (a) or blur with the background~(f).
Furthermore, it is noticeable that the distribution of the particles is essential. Thus, in (c) and (d), there are only particles that are organized in a row or in groups, which is not the case for large particles.

Now the question is: How to extract features from a grayscale image, namely an array of integers between $0$ (black) and $255$ (white), such that the features above are captured algorithmically and a classifier is able to separate the two classes mentioned above?

\begin{figure}[!htbp]
	\centering
	\subfloat[Usage of the intensity average of each column of an image as a feature regarding particle size and distribution for the case of videos including parallel-stripe domain pattern.]{\includegraphics[width=6.4cm]{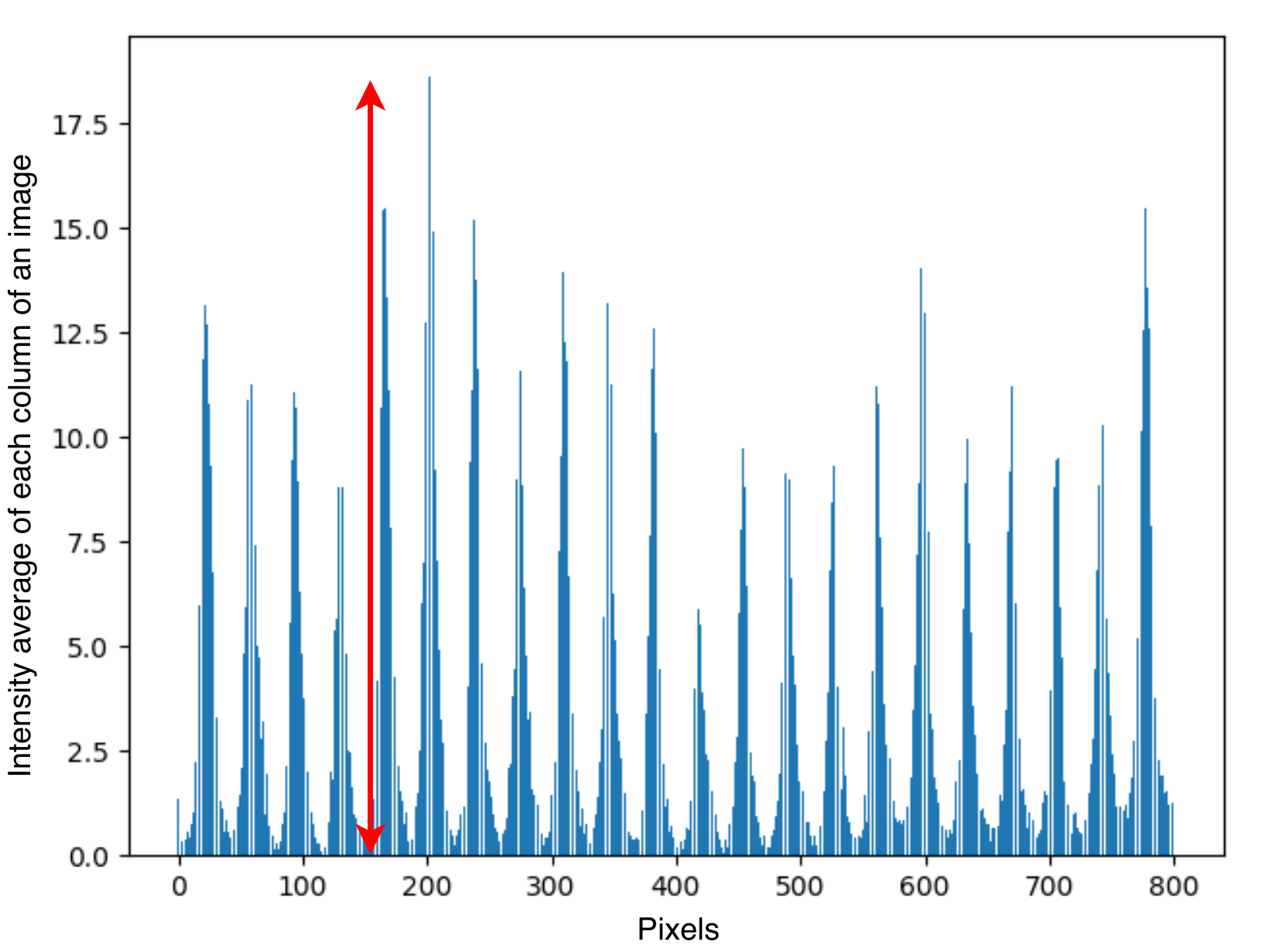}}\qquad
	\subfloat[The corresponding intensity average of each row of an image within videos including parallel-stripe domain pattern.]{\includegraphics[width=6.4cm]{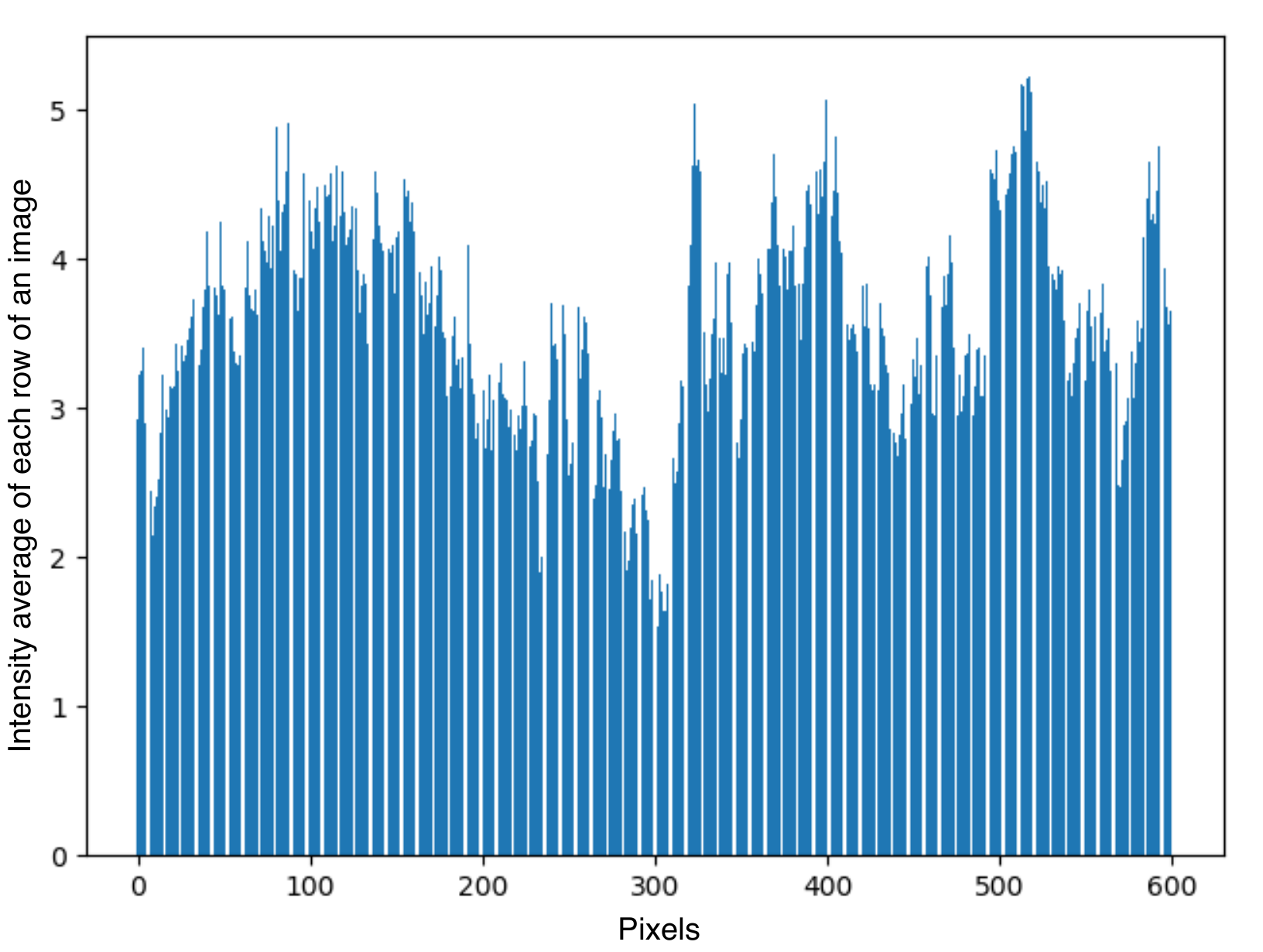}}\qquad
	\caption{Features extracted from the intensity average of each column/row within a parallel-stripe domain pattern, cf. Fig.~\ref{fig:examplesSelection}~(c).}
	\label{fig:distributionParticles}
\end{figure}

To estimate the distribution of the particles, it is possible to compute the average grayscale value over all rows and columns in one single frame. Since images in the videos are in grayscale and $0$ represents black ($255$ white, respectively), an estimate can be made regarding the number of particles. In the following, the frame is regarded as a two-dimensional array and is thus referred to as \textit{rows} and \textit{columns}. In other words, at a high particle density in a frame after background subtraction, the average value of the corresponding rows or columns will be correspondingly high (because the particles are bright) and vice versa. 
In order to express the particle density within a video as a feature, reference is made to the individual cases in Fig.~\ref{fig:examplesSelection}. In the case of (c), it is clearly visible that the particles follow a pattern. Thus, it can be seen that the particle density, if divided into several columns, is high in some areas and low in others, cf. Fig.~\ref{fig:distributionParticles} (a). For the same (in this case parallel-stripe) pattern, the particle density with regard to several rows of the image is relatively uniform, cf. Fig.~\ref{fig:distributionParticles} (b). In case (d), the distribution of the particles regarding the rows should be similar to the distribution in Fig.~\ref{fig:distributionParticles} (a), for instance. In some cases, this information can be used to estimate a video's particle density.
Fig.~\ref{fig:distributionParticles} shows the intensity average of each column of the image using the application case of a parallel-stripe domain pattern. The peaks indicate the distribution of the particles in the image. High peaks indicate that many particles are in these areas and vice versa. The \textit{range}, cf. red-colored arrow in (a), of the respective two histograms in Fig.~\ref{fig:distributionParticles} is therefore useful to determine an initial indication of the distribution of the particles within the image. 
Furthermore, it may be relevant to consider the \textit{number of pixel values above a certain threshold}. This number can provide information about the underlying distribution of particles in the videos as well.

The size of the particle images can be expressed using a radius or a diameter. A radius can be obtained using \textit{circular Hough transform}, which has already been used to fill particles in Sec.~\ref{impl:preprocessing}. \textit{Averaging the radii} of all detected circles provides a suitable feature regarding the estimation of a particle size as a first estimate.
Another method to approximate a particles size in some cases is, e.g., using the average of each row or column in Fig.~\ref{fig:distributionParticles}. By computing the \textit{peaks width} of the data, the particle sizes can be estimated as well in some cases.

These selected features are used to train (and afterward query) an \textit{ensemble random forest classifier} \cite{BreimanLeo2001}, which can automatically determine the selection of the appropriate preprocessing sequence after successful training.

\subsection{Regression} \label{ap:regression}

For particle localization, two parameters are required, namely the \textit{particle size} and a \textit{minimum mass}. To provide these parameters in an automated manner, these have to be estimated. This, in turn, can be done by using regression models, which for each video input provide values for the necessary parameters. In order to train the regression models, again, features from the video frames are required. To better qualify the particular features, which are required for training, two distinct \textit{logistic regression models} were created for the respective two parameters.

Both requested parameters clearly depend on the size of the particles, cf. Tab.~\ref{tab:trackpy}. The distribution of the particles in the image could be of importance as it allows for inferences about the size of the particles in some application cases, cf. Sec.~\ref{ap:classification}. What is striking is that these features have already been used for the classification problem in the previous section. Therefore, the features computed for training the classifier in the previous section may likewise be used for regression model training.

\unappendix

\bibliographystyle{elsarticle-num}
\bibliography{Literature.bib}

\end{document}